\documentclass[%
reprint,
groupedaddress,
bibnotes,
amsmath,amssymb,
aps,
]{revtex4-2}

\usepackage{titlesec}

\renewcommand\thesection{\arabic{section}}
\renewcommand\thesubsection{\thesection.\arabic{subsection}}

\titleformat{\section}{\normalfont\Large\bfseries}{\thesection}{1em}{}
\titleformat{\subsection}{\normalfont\large\bfseries}{\thesubsection}{1em}{}

\makeatletter
\renewcommand\p@subsection{}
\makeatother

\usepackage{graphicx}
\usepackage{dcolumn}
\usepackage{bm}

\begin{document}

\title{Designing a braille reader using the \\ snap buckling of bistable magnetic shells}

\author{Arefeh Abbasi$^1$}
\author{Tian Chen$^2$}
\author{Bastien F.G. Aymon$^1$}
\author{Pedro M. Reis$^{1*}$}

\affiliation{$^1$Flexible Structures Laboratory, Institute of Mechanical Engineering,  \\ \'{E}cole Polytechnique F\'{e}d\'{e}rale de Lausanne (EPFL), 1015 Lausanne, Switzerland \\
$^2$Architected Intelligent Matter Laboratory, Department of Mechanical Engineering, University of Houston, Houston, TX, 77024, U.S.A. \\ $^*$Email: pedro.reis@epfl.ch
}










\begin{abstract}
A design concept is introduced for the building block, a \textit{dot}, of programmable braille readers utilizing bistable shell buckling, magnetic actuation, and pneumatic loading. The design process is guided by Finite Element simulations, which are initially validated through precision experiments conducted on a scaled-up, single-shell model system. Then, the simulations are leveraged to systematically explore the design space, adhering to the standardized geometric and physical specifications of braille systems. The findings demonstrate the feasibility of selecting design parameters that satisfy both geometric requirements and blocking forces under moderate magnetic fields, facilitated by pneumatic loading to switch between the two stable states. The advantages of the proposed design include the reversible bistability of the actuators and fast state-switching via a transient magnetic field. While the study is focused on experimentally validated numerical simulations, several manufacturing challenges that need to be resolved for future physical implementations are identified. \\

\end{abstract}

\maketitle
\newpage

\section{Introduction}
\label{Introduction}

Worldwide, 285 million people experience visual impairments, including 39 million living with blindness~\cite{pascolini2012global}. These impairments present challenges in navigating and interacting with the world, influencing various aspects of daily life, such as access to printed or digital content. Braille code, a tactile writing system, facilitates these interactions, mapping symbols (\textit{e.g.}, letters, numbers, and punctuation) into arrays of cells, each comprising a 3$\times$2 matrix of dots. Each dot can independently be raised or flat, and words are then formed by assembling a series of such cells. Braille users typically read by tracing their fingertips across rows of these cells, whose dimensions are optimized to allow the index finger pad to cover the entire cell and discern each dot. While static embossed paper is the traditional medium for braille, modern assisting devices, such as refreshable braille displays (RBDs), enable dynamic reading and writing ~\cite{leonardis2017survey,jimenez2009biography}. 
RBDs provide access to written content using arrays of morphable physical dots, whose configuration  adjusts dynamically to represent different sequences of braille symbols in accordance with international braille standards~\cite{lee2005micromachined,wu2007portable,runyan2011seeking, leonardis2017survey}. 

In recent years, technological advancements have stimulated the development of RBD devices driven by a variety of actuation mechanisms, including piezoelectrics~\cite{ smithmaitrie2009analysis,smithmaitrie2007touching, velazquez2008tactile, hernandez2009characterization}, electromagnetics~\cite{simeonov2014graphical,bettelani2020design,balabozov2014computer}, or thermopneumatics~\cite{vidal2005thermopneumatic}. These devices often use advanced materials such as electroactive polymers~\cite{chakraborti2012compact, ren2014new, frediani2018enabling,qiu2018refreshable}, shape memory alloys~\cite{makishi2001batch,velazquez2007toward, besse2017flexible}, or dielectric elastomers ~\cite{koo2008development,matysek2009dielectric}. Piezoelectric actuators have been favored for commercial RBDs due to their fast refresh rates, low power consumption, and reliability, albeit at a relatively high cost~\cite{leonardis2017survey, smithmaitrie2009analysis,smithmaitrie2007touching}. Electromagnetic linear actuators tend to have a low ratio between output force and operating velocity, requiring complex packaging and a considerable force to hold a raised dot~\cite{karastoyanov2014electromagnetic, yatchev2011static}. RBD devices utilizing shape memory alloys require intricate heating and cooling processes, posing practical implementation challenges~\cite{velazquez2008tactile, besse2017flexible, makishi2001batch}. Dielectric elastomers have been gaining traction for lightweight tactile displays, offering high actuator density and a wide range of motion, with performance comparable to previous technologies, but in a more compact form~\cite{koo2008development,matysek2009dielectric}. However, their required large driving voltages can be impractical for some applications.

To date, most technological solutions for RBDs cannot offer sufficiently high-quality performance, especially regarding the balance between fast shape-changing dynamics and low power consumption~\cite{king2010perceptual}. Additionally, the mainstream adoption of advanced tactile displays is hindered by the lack of compact, large-area actuator arrays that can stimulate multiple sensory receptors while adhering to high user-safety standards. Existing solutions tend to be costly and require complex manufacturing processes. Despite ongoing efforts~\cite{ishizuka2015mems}, designing RBD devices that are simple, compact, low-cost, large-scale, user-friendly, and reliable remains a formidable challenge.

Here, we propose a novel design concept for braille dots, the building block of braille readers. Our design leverages the buckling and bistability of thin shells  fabricated from hard magneto-rheological elastomers (h-MREs). Inspired by the popular `\textit{pop it}' toy~\cite{popit}, these shells can be reversibly set in a convex or concave state (\textbf{Figure~\ref{Fig1}}a). Each of these shells (dots) can then be arranged in a 3$\times$2 matrix and programmed, on-demand, to form a braille symbol. The dots have independent writing and reading phases under magnetic and mechanical loading, respectively. During writing, a transient external magnetic field can induce snap-through buckling to  transition the shell between its two stable states: from ON (bump) to OFF (dimple) or vice versa. For reading, shells in the ON state must sustain a  blocking force, in reaction to the finger indentation, without snapping to the OFF state. Throughout, our mechanics-based design process is centered on Finite Element Method (FEM) simulations. We first validate these simulations against precision experiments on a scaled-up (centimeter-scale) physical model of a braille dot. Then, we study dots at their actual scale, ensuring adherence to standard braille specifications~\cite{runyan2011seeking}, with a special focus on their geometry, elastic response, and actuation. Although the primary driver of actuation is an external magnetic field, it is supplemented with a transient pneumatic loading to aid in widening the design space. Our numerical exploration of the design parameters allows us to identify the regions that meet the various design constraints, making a step toward a new class of programmable braille displays.

\section{Problem definition: braille reader design concept}
\label{Braille_Design}

The specifications for braille cells and dots are standardized by the World Blind Union~\cite{runyan2011seeking}, and the relevant parameters are required to lie within the following ranges: $\ell_\mathrm{d} \in [2.3,\,2.5]\,$mm for the dot-to-dot spacing, $\ell_\mathrm{c} \in[6,\,7]\,$mm for the distance between two distinct cells, and $\ell_\mathrm{l} \in[10,\,11]\,$mm for the distance between two lines of words. Furthermore, each raised dot must feature a quasi-hemispherical cap with base diameter $D\in[1.4,\,1.6]\,$mm and height $ h\in[0.4,\,0.9]\,$mm~\cite{LibraryofCongress2014}. Finally, to sustain the normal indentation force applied by the index finger during reading, each dot must be able to withstand a minimum blocking force of $F>50$~mN~\cite{runyan2011seeking}. 

\begin{figure}[h!]
        \centering
        \includegraphics[width=1\columnwidth]{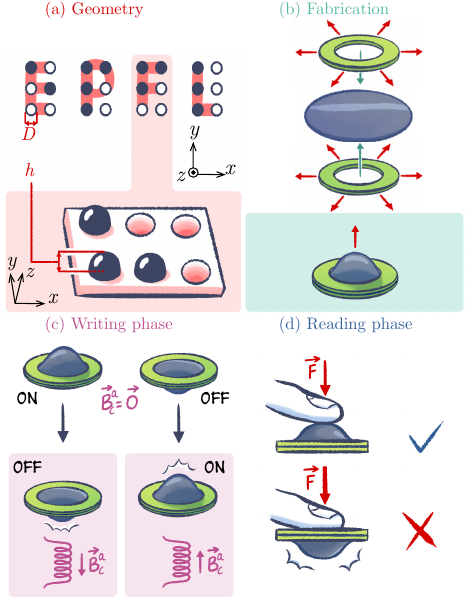}
        \caption{\textbf{Design, fabrication, and operation of a bistable Braille dot.} (\textbf{a}) \textit{Geometry:} A word is formed by assembling a series of braille cells, each comprising $3\times2$ dots. ``EPFL" is shown as an example. (\textbf{b}) \textit{Fabrication:} A bistable shell is fabricated by sandwiching a circular h-MRE plate between two radially pre-stretched boundary annuli. This pre-stretch is then released to buckle the plate into a shell.
        (\textbf{c}) \textit{Writing phase:} an external magnetic field, $\mathbf{B}^\mathrm{a}$, sets each shell in one of its two stable states, either ON (bump) or OFF (dimple).
        (\textbf{d}) \textit{Reading phase:} an index finger applies an indentation force, $F$ on each dot.
        }
        \label{Fig1}
\end{figure}

Our objective is to design a programmable braille dot that adheres to the aforementioned braille specification. We consider a  bistable shell  clamped at its base (Figure~\ref{Fig1}a, b). The analysis is segmented into two phases: ``\textit{Writing}'' (Figure~\ref{Fig1}c) and ``\textit{Reading}'' (Figure~\ref{Fig1}d). For writing, the ON-OFF switching is done via magnetic actuation. Concurrently with this phase, we temporarily depressurize the shell to lower the energy barrier required for buckling. By contrast, during the reading phase, the shell is pressurized to increase its rigidity. We seek to identify the key design variables and protocols required for the fabrication and operation of our system. Next, we describe the geometric considerations and the two operational phases.

The geometry of our model braille-dot (Figure~\ref{Fig1}a) comprises a  shell of diameter $D$=1.45\,mm and height $h$=0.48\,mm, in accordance with braille standards. This shell is fabricated by the buckling of a radially compressed circular plate (thickness $t$) made of h-MRE~\cite{yan2022reduced} when the in-plane pre-stretch, $\lambda$, of two boundary annuli is released (Figure~\ref{Fig1}b), as detailed in Section~\ref{Experiments}. One first goal of the design is to select appropriate values of $t$ and $\lambda$ that, upon buckling of the plate, yield a shell with the target value of $h$, satisfying braille requirements.

For the writing phase (Figure~\ref{Fig1}c), we will characterize the snap buckling of the shells under loading by a uniform magnetic field, $\textbf{B}^\mathrm{a}_{c}$, to switch between their two stable states. We assume that each dot, which would eventually form the 3$\times$2 cell, can be actuated independently. For the present study, we restrict our focus to the operation of a single dot. The goal of  this design phase is to identify the critical magnetic field, $B^\mathrm{a}_\mathrm{c}$, required for snapping under the limitation set by upper-bound of the linear regime of the \textbf{B}-\textbf{H} hysteresis curve for the h-MRE material~\cite{zhao2019mechanics} (additional details are provided in Section~\ref{Experiments}). Subsequently, we aim to determine the corresponding geometric and fabrication parameters, $t$, and $\lambda$, that yield the desired snap-buckling characteristics.

For the reading phase (Figure~\ref{Fig1}d), the braille dots must be designed such that the user can tactilely discern the dots without altering their state. The challenge lies in ensuring that a dot in the ON state can sustain the indentation force mentioned above ($F\geq50$\,mN) without snapping to the OFF state, thereby inadvertently erasing the braille pattern. This design phase targets the determination of optimal geometric and fabrication parameters for the dot, specifically its thickness $t$ and pre-stretch $\lambda$, to meet this requirement.
    
The constraints associated with the fabrication protocol and the ensuing dot geometry, coupled with the requirements for the reading and writing phases, underscore the intricacies involved with designing our braille dot. We aim to identify the feasible design parameter space of the system (specifically, $t$ and $\lambda$) that satisfies the constraints on $h$, $F$, and $B^\mathrm{a}_\mathrm{c}$. This design exploration will be performed solely using FEM simulations, which will be initially validated against experiments in a scaled-up system.

\section {Results and Discussion}
\subsection {Validation of the FEM simulations against experiments}
\label{Validation}

We first validate the FEM simulations (technical details are provided in Section~\ref{FEM}) against experiments (see Section~\ref{Experiments}) on the scaled-up model system, considering the results from the geometry characterization of the fabricated shells, as well as from the reading and writing experiments.

\begin{figure}[ht!]
        \centering
        \includegraphics[width=1\columnwidth]{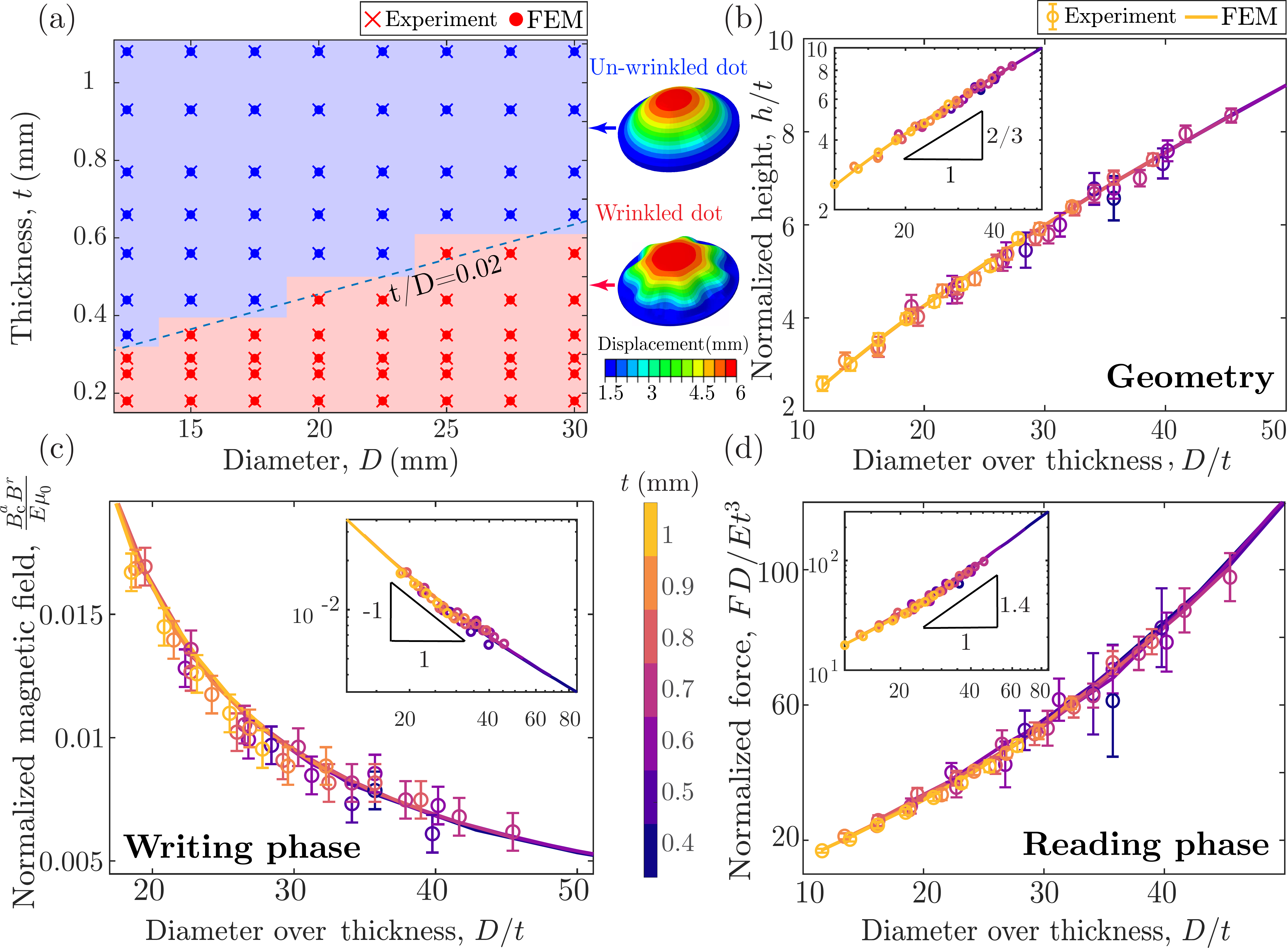}
        \caption{\textbf{Validation of FEM simulations against the scaled-up experiments.} (\textbf{a}) Phase diagram in the ($t,\,D$) parameter space. The experiments and simulations correspond to the cross and circle symbols, respectively. The smooth and wrinkled shells are represented by the blue and red symbols, respectively; the empirical phase boundary between the two is represented by the dashed line. The normalized (\textbf{b}) shell height, $h/t$, (\textbf{c}) critical magnetic field, $B^\mathrm{a}_\mathrm{c}B^\mathrm{r}/(E \mu_0)$, and (\textbf{d}) blocking force, $FD/(Et^3)$ are plotted as functions of the normalized base diameter of the shell, $D/t$, for 10 different values of thickness, $t$. The error bars of the experimental data represent the standard deviation of 6 independent measurements on the same specimen. The different values of $t$ are color coded (see adjacent color bar). The solid lines and data symbols correspond to FEM and experiments, respectively. The insets show the log-log plots of the data.  Throughout, the fabrication pre-stretch is $\lambda=0.1$.}
        \label{Fig2}
\end{figure}

During sample fabrication, the buckling of the plate, which yields a shell, may produce undesirable wrinkling patterns~\cite{davidovitch2011prototypical}. To act as braille dots, ideal shells should be smooth (\textit{i.e.}, free of these wrinkles). Toward identifying the design space for these ideal shells, \textbf{Figure~\ref{Fig2}}(a) presents a phase diagram of the thickness-diameter parameter space ($t,\,D$) for representative shells fabricated with a pre-stretch of  $\lambda=0.1$ (see Section~\ref{Experiments}, and~\ref{FEM}). There is excellent agreement between experiments (crosses) and FEM (circles), serving as a first step in validating the simulations. Wrinkling is observed for higher values of the slenderness ratio $D/t$. For the chosen pre-stretch ($\lambda=0.1$), the empirical phase boundary between ideal and wrinkled shells is $t\approx 0.02 D$ (dashed line).

First, focusing on the geometry characterization, Figure~\ref{Fig2}(b) plots the normalized height, $h/t$, of the smooth shells (blue region in Figure~\ref{Fig2}a) versus $D/t$. The FEM simulations (lines) and experimental data (symbols) are obtained with a fabrication pre-stretch of $\lambda=0.1$ and thickness in the range $t\in[0.18,\,1.08]\,$mm (see Section~\ref{Experiments}). The data collapses onto a single curve, with excellent agreement between experiments and FEM, thereby validating the FEM for the shell fabrication. Moreover, the inset of Figure~\ref{Fig2}(b) shows that the relation between $h$ and $D$ follows a power law, consistently with the prediction from Föppl–Von Kármán theory for the buckling of a circular plate under in-plane radial compression: $h/t \sim \left(D/t\right)^{2/3}$~\cite{timoshenko1959theory}.  

Toward validating the writing-phase simulations, in Figure~\ref{Fig2}(c), we plot the normalized magnetic field required for snapping, $B^\mathrm{a}_\mathrm{c} B^\mathrm{r}/(E\mu_\mathrm{0})$, as a function of $D/t$. Naturally, increasingly slender shells require a lower magnetic field for snapping, and the data collapse into a single curve, exhibiting a power law with an exponent $\approx -1$. This scaling originates from the balance between magnetic and elastic energies investigated in our previous work~\cite{yan2021magneto, pezzulla2022geometrically}, which suggests $B^\mathrm{a}_\mathrm{c} B^\mathrm{r}/(E\mu_\mathrm{0}) \sim (D/t)^{-1}$; a prediction in agreement with our present data (inset of Figure~\ref{Fig2}c). 

In Figure~\ref{Fig2}(d), we present the results for the reading, plotting the normalized blocking force, $FD/(Et^3)$ (required to snap the shell), as a function of $D/t$. Again the experiments (symbols) are in excellent agreement with the FEM (lines). The data is consistent with a power law with exponent $\approx1.5$, which can be rationalized using well-established results for the indentation of a spherical shell by a flat plate, causing mirror buckling (to produce an inverted cap) of the shell~\cite{landau1986theory,  pogorelov_bendings_1988, audoly2010elasticity,vaziri2009mechanics}. Balancing  the stretching and bending energies of the shell, the dimensionless indentation force is expected to scale  as $FD/(Et^3) \sim (D/t)^{3/2}$, a prediction that is consistent in our data (inset of Figure~\ref{Fig2}d). For completeness, the dimensional version of the plots shown in Figure~\ref{Fig2}(b,\,c,\,d) are provided in Supplementary Information.

Overall, we found excellent agreement between the experiments and the FEM simulations of the scaled-up model system for the shell-fabrication protocol and their geometric characterization, as well as for the reading and writing phases. 


\subsection{Design of braille dots at the real scale.}
\label{Micro Braille}

We shift our attention from the scaled-up model system to investigate the design of the real-scale braille dots. Leveraging the FEM simulations validated above, and following the protocol details in Section~\ref{FEM}, we explore the design space for braille dots and determine their optimal fabrication and operational conditions. Each dot must conform to the specifications laid out in Section~\ref{Braille_Design}. Our objective is to determine the optimal ranges for the key geometrical parameters (thickness, $t$, and fabrication pre-stretch, $\lambda$) that simultaneously meet the acceptable design constraints for dot height (obtained from  fabrication) and meet feasible operational conditions for the writing phase (critical magnetic field for actuation), and reading phase (blocking force). Finally, we will identify the intersecting region of these three design sub-spaces.

 \begin{figure}[b!]
        \centering
        \includegraphics[width=1\columnwidth]{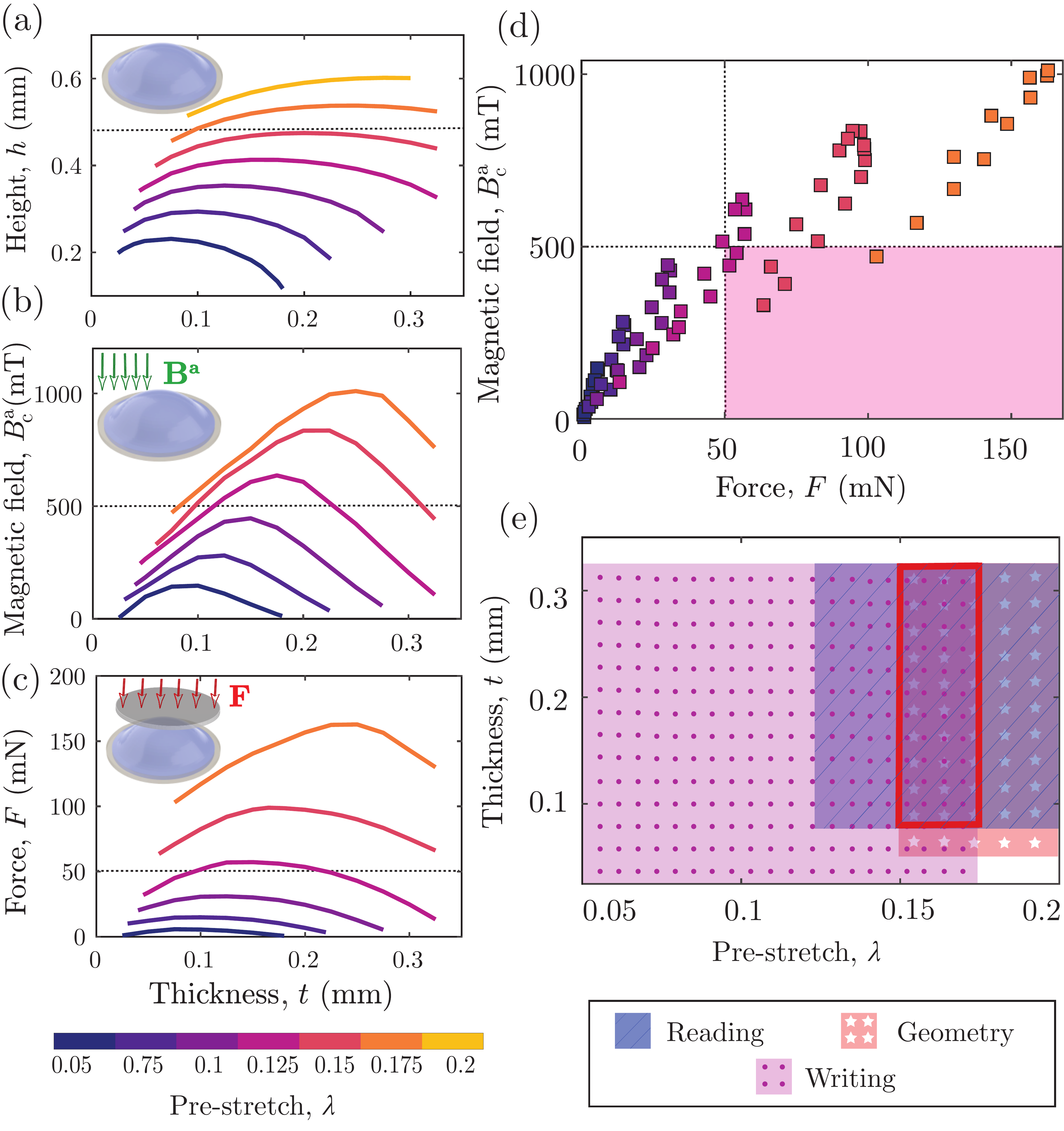}
        \caption{\textbf{Design of the real-scale braille dot.} The (\textbf{a}) dot height, $h$, (\textbf{b}) critical magnetic field, $B^\mathrm{a}_\mathrm{c}$, and (\textbf{c}) blocking force, $F$, are all plotted versus thickness, $t$, at different levels of pre-stretching, $\lambda$=$[0.05,0.2]$ (increments of 0.05). The horizontal dashed lines represent the design constraints of the corresponding quantities. (\textbf{d}) Phase diagram in the ($B^\mathrm{a}_\mathrm{c}$, $F$) parameter space, with the desired shaded region. (\textbf{e}) Phase diagram in the ($t$, $\lambda$) parameter space for the geometry step and reading and writing phases, with the feasible design space indicated by the rectangle region. All results were obtained  from FEM simulations.}
        \label{Fig3}
    \end{figure}

\begin{figure*}[ht!]
        \centering
        \includegraphics[width=1\textwidth]{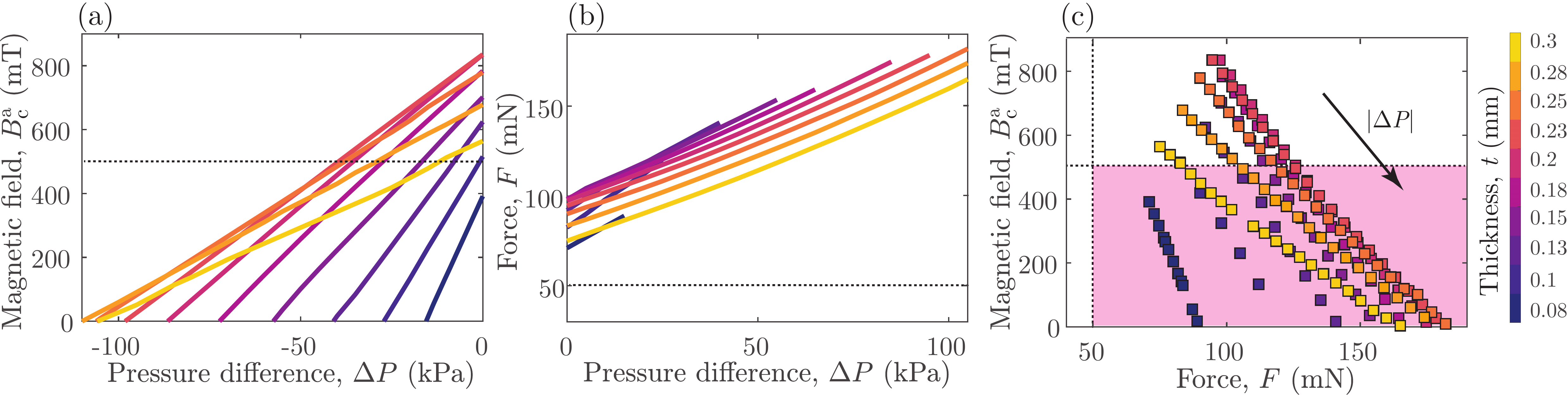}
        \caption{\textbf{Improved design of real-scale braille dots with pneumatic system.} (\textbf{a})  Blocking force, $F$, and (\textbf{b}) magnetic field, $B^\mathrm{a}_\mathrm{c}$, plotted as a function of the pressure difference, $\Delta P$. The thickness was varied in the range $t=[0.075-0.3]\,$mm  (in increments of 0.025\,mm) for a specific pre-stretch $\lambda=0.15$. The horizontal dashed lines represent the limiting bounds imposed by the  braille standards. (\textbf{c}) Phase diagram in the ($B^\mathrm{a}_\mathrm{c}$, $F$) parameter space with the desired regions of viable parameters (shaded region), for different levels of pressure difference, $0<|\Delta P|<120$~kPa. All results were obtained from FEM simulations.}
        \label{Fig4}
\end{figure*}

First, we characterize the geometry of the dots obtained from the fabrication step of the simulations.
In \textbf{Figure~\ref{Fig3}}(a), we plot the dot height, $h$, versus thickness, $t$, for different levels of $\lambda$. The color map represents the various levels of $\lambda$, whose range is specified in Section~\ref{FEM}. The resulting $h(t)$ curves exhibit a non-monotonic trend, with overall values (including the maximum of the curves) that increase with $\lambda$. This non-monotonic behavior arises because for either very thick or thin plates, the pre-stretch release (compression) leads to planar (radial) contraction rather than increasing the out-of-plane deformation of the buckled plate (shell). The horizontal dashed line represents the minimum dot height, $h\geq 0.48\,mm$, required by braille specifications. Thus, to satisfy this requirement, we find that the feasible range for the fabrication pre-stretch is $\lambda \geq 0.15$, and the viable thickness range is $t \in [0.05, 0.325]\,$mm.

Next, we consider the writing phase, which enables the braille dot (magnetic shell) to switch between its two stable stages. In Figure~\ref{Fig3}(b), we plot the critical amplitude of the magnetic field, $B^\mathrm{a}_\mathrm{c}$, required to snap the dot as a function of $t$, for different fabrication pre-stretches. We observe that, $B^\mathrm{a}_\mathrm{c}$ increases with $\lambda$ (and thus for taller dots; cf. Figure~\ref{Fig3}a), also with a non-monotonic dependence on $t$. To prevent the demagnetization of the shell due to high magnetic fields~\cite{zhao2019mechanics}, we established the upper limit ${B^\mathrm{a}_\mathrm{c}} \leq$500\,mT, represented by the horizontal dashed line in Figure~\ref{Fig3}b). Consequently, for actuation within this magnetic-field limit, the allowed parameters are in the ranges $t \in [0.025,\, 0.325]\,$mm, and $\lambda \in [5,\,17.5]$.

Finally, we turn our attention to the reading phase. In Figure~\ref{Fig3}(c), we plot the blocking force, $F$, as a function of $t$, using the same ranges of the other parameters which are specified in Section~\ref{FEM}. Increasing $\lambda$ leads to an overall increase of the $F(t)$ curves, much like the writing phase, which is also non-monotonic. Combining these results with the $h(t)$ data in Figure~\ref{Fig3}(a) implies that taller dots require a higher indentation force for inversion, presumably due to their geometry-induced rigidity~\cite{Lazarus2012,Vella2012}. According to braille specifications, the blocking force of the dot must be at least $F \geq 50\,$mN (horizontal dashed line in Figure~\ref{Fig3}c), which is limited by the potential snapping of the shell due to the touch by a fingertip. Under this constraint, we determine that the ranges of feasible parameters for this reading phase are  $t \in [0.075,\,0.325]\,$mm, and $\lambda \geq 12.5$.

Combining the above results for the viable ranges of the parameter space $(t,\,\lambda)$ dictated by the geometry characterization, writing, and reading phases, we present the intersection of these three design phases in Figure~\ref{Fig3}(d). At each level of $\lambda$ (colored symbols), we plot $B^\mathrm{a}_\mathrm{c}$ as a function of $F$ for all the thickness values. An overall correlation emerges between $B^\mathrm{a}_\mathrm{c}$ and $F$. In the plot, the design constraints on $B^\mathrm{a}_\mathrm{c}$ and $F$ mentioned above are represented by the shaded region, which only intersects with a few of the explored designs ($\lambda \geq 0.125$). The design constraints require a sufficiently high blocking force while ensuring a sufficiently low magnetic field, a trade-off that is challenging to achieve in our system.

In Figure~\ref{Fig3}(e), we present an alternative version of the overlap of all of the design constraints explored above, now in the final target design parameter space $(t,\,\lambda)$. Each separate shaded/textured region relates to the individual viable bounds obtained above for the geometric characterization, writing, and reading phases. Meeting all the constraints and ensuring braille standards requires an overlap of these three regions; \textit{i.e.},  the domain enclosed by the dashed rectangle with $t \in [0.1, 0.325]\,$mm and $\lambda \in [0.15, 0.175]$. 


\subsection{Design improvement using a pneumatic system}
\label{pressure}

The feasible design space identified from the results in the previous section is rather limited, making a flexible fabrication process challenging. The design requires the shell to be able to snap during the writing phase yet remain resistant to snapping during the reading phase. To address these conflicting limitations without altering the geometry, we propose the incorporation of an additional pneumatic loading system. This pneumatic component modulates the energy barrier for buckling in both the reading and writing phases; the pressure difference, $\Delta P$, between the inside and outside of the dot is positive for the reading phase and negative for the writing phase. The implementation details of this pneumatic loading in the FEM simulations are provided in Section~\ref{FEM}. Hereon, we focus on shells fabricated with a pre-stretch of $\lambda=0.15$, which was deemed practical from the parameter exploration presented above.

For the writing phase, the dot is depressurized to reduce the critical magnetic field $B^\mathrm{a}_\mathrm{c}$ required for snapping, thereby facilitating the switching of the dot. In \textbf{Figure~\ref{Fig4}}(a), we plot $B^\mathrm{a}_\mathrm{c}$ versus $\Delta P$, for different values of $t$. The results show that applying pneumatic loading substantially reduces the magnetic field $B^\mathrm{a}_\mathrm{c}$. The $B^\mathrm{a}_\mathrm{c}(\Delta P)$ curves are linear, with a slope that varies with $t$. For example, depending on $t$, a dot depressurized by $\Delta P \lesssim -60\,$kPa can lower the critical magnetic field for snapping by as much as 50\%, compared to the zero-pressure case. The horizontal dashed line represents the maximum acceptable magnetic field. 

For the reading phase, in contrast to the writing phase, the dot is pressurized to enhance the blocking force, $F$, and better resist indentation. The dot is first pressurized and then loaded at its pole. In Figure~\ref{Fig4}(b), we present the dependence of the blocking force $F$ on the  applied $\Delta P$ for different thickness values, $t$ (see adjacent color bar). In the explored range of parameters, the $F(\Delta P)$ response is linear, as expected from previous work~\cite{Lazarus2012, Vella2012}, and no buckling occurs. For example, the resistance of the dot to indentation force can be increased by up to 100$\%$ (at $t=0.2\,$mm) compared to the zero-pressure case. We find that all curves are now well above the $50\,$mN limit imposed by braille standards.

Finally, in Figure~\ref{Fig4}(c), we combine data from both phases with pneumatic loading (Figure~\ref{Fig4}a and~\ref{Fig4}b) and plot $B^\mathrm{a}_\mathrm{c}$ versus $F$. The shaded region indicates the viable range of parameter space; with a pneumatic load of $|\Delta P| \geq 40\,$kPa, we achieve successful reading and writing operations across the full thickness range. The limit values of $|\Delta P|$ could be further tuned by varying the fabrication pre-stretch $\lambda$, but we leave a more systematic exploration for future work. Our results demonstrate that using this additional pneumatic component, the design space of the system is significantly expanded compared to the zero-pressure case explored in Section~\ref{Micro Braille}.

\section{Conclusion}
\label{Conclusions}

We have proposed a new design concept for a reversibly switchable braille dot as a building block for refreshable braille displays. The proposed mechanism uses bistable magnetic shells that can be snapped on-demand under an external magnetic field (writing phase) while resisting buckling due to the indentation by a fingertip (reading phase). An additional pressure-loading component expands the available design space without modifying the dot geometry. First, we performed experiments on a scaled-up model system to validate FEM simulations. These simulations were then leveraged to systematically explore the design space at realistic scales while meeting braille standards (geometry and tactile sensitivity) with reasonable magnetic field strengths and temporary pneumatic loading. Our design boasts several advantages over existing solutions, including bistability for self-stabilization, as well as fast state-switching and pattern refreshment. This switching can be triggered by a transient magnetic field rather than a continuous energy input. Finally, a constant pneumatic input for the whole actuator enables the tuning of the power input of the system. 

While our design introduces promising advances, it is not without potential limitations. Its complexity calls for advanced manufacturing and assembly techniques. Miniaturized solenoids to generate the required magnetic field under each dot would need to be developed. Furthermore, incorporating a hybrid magnetic and pneumatic system could pose challenges in terms of size and power, particularly for portable or battery-operated devices. Despite these potential obstacles, we anticipate that future research and physical implementation of this concept could make it possible to build a new class of compact, user-friendly, and cost-effective braille readers.


\section{Methods Section}
\subsection{Experiments with the scaled-up system}
\label{Experiments}

In this section, we detail the fabrication protocol, geometric characterization, and testing (writing and reading phases) for the experiments on our scaled-up model system. The data obtained from these experiments serve to validate the FEM simulations in Section~\ref{Validation}.

\begin{figure}[t!]
        \centering
        \includegraphics[width=1\columnwidth]{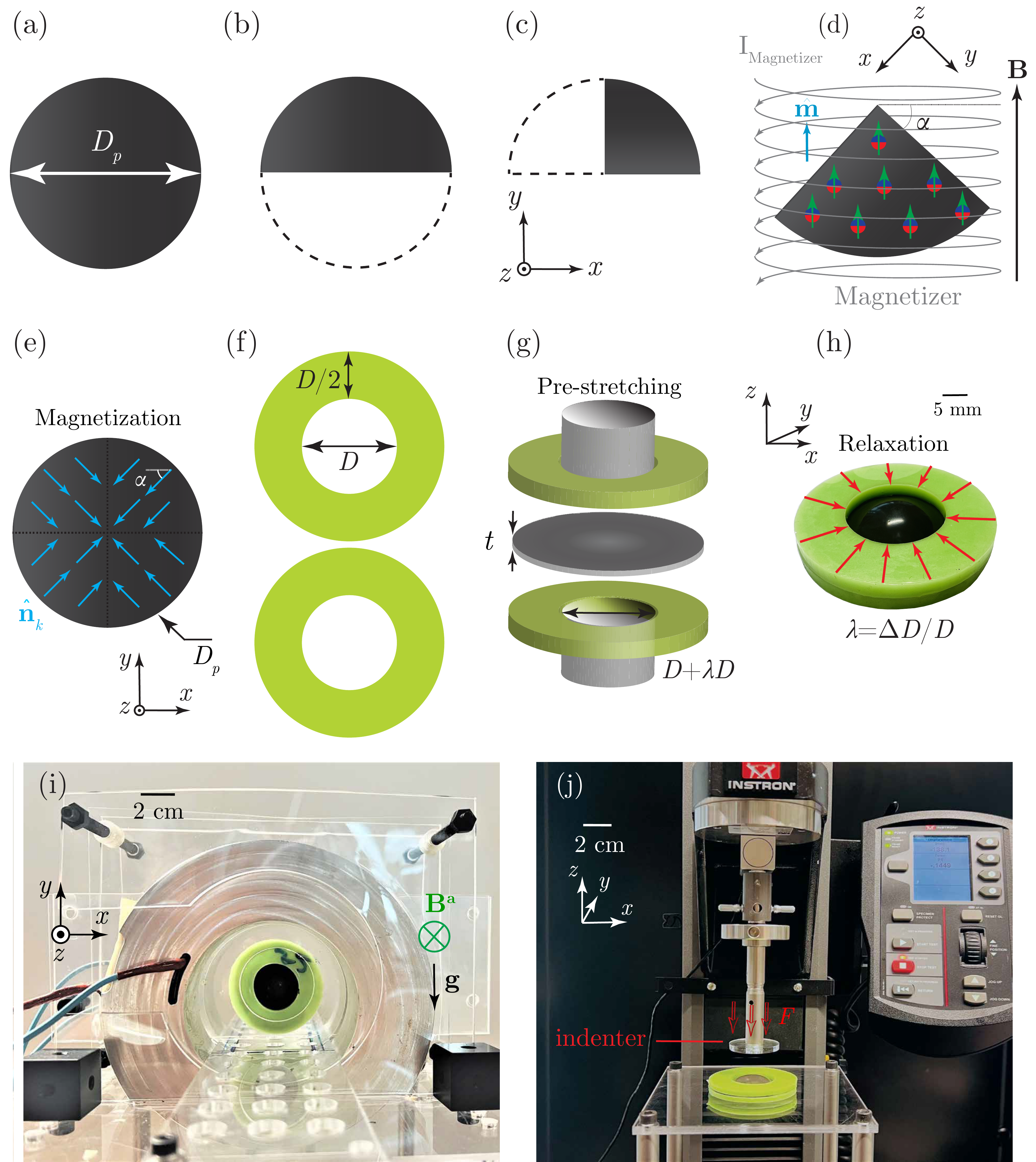}
        \caption{\textbf{Fabrication and experimental apparatus for the scaled-up model.} 
        (\textbf{a}) A circular magnetic plate (diameter of $D_\mathrm{p}$) is first (\textbf{b}) folded along the $x$-axis and then (\textbf{c}) along the $y$-axis to form a quarter-circle shape. (\textbf{d}) The folded plate is inserted into  the pulse magnetizer at an angle $\alpha=45^{\circ}$. 
        The magnetizer generates a strong axial magnetic field, \textbf{B}, which induces a magnetic moment, $\mathbf{m}$. (\textbf{e}) The circular h-MRE plate was magnetized while folded, yielding the magnetization profile described by Equation~(\ref{eq:magnetization}). (\textbf{f}) Two VPS annuli serve to constrain the plate boundary. (\textbf{g})  Cylinders pre-stretched the annuli, which  sandwich the h-MRE plate. (\textbf{h}) Upon release of the pre-stretch, the plate buckles to form a shell. (\textbf{i}) Photograph of the  apparatus for the \textit{writing} experiments. The shell is placed between two Helmholtz coils and loaded by a uniform magnetic field  along $\hat{\mathbf{z}}$. (\textbf{j}) Photograph of the apparatus for the \textit{reading} experiments; a plate indents the shell, and the blocking force $F$ is measured.
        }
        \label{Fig5}
\end{figure}

\textit{Fabrication:} We have followed an established experimental procedure to prepare the h-MRE material used to fabricate our specimens~\cite{yan2021comprehensive, abbasi2022snap, yan2022reduced}. First, we  mixed Vinylpolysiloxane  (VPS-32, Elite Double, Zhermack) with NdPrFeB particles (MQFP-15-7-20065-089, Magnequench), with volume fraction $c_v$=18.7$\%$. Then, an automated film applicator (ZAA 2300, Zehntner) spread the VPS-NdPrFeB mixture into a thin film, which, upon  curing, yielded a thin elastic plate. By modulating the gap height of the film applicator, we fabricated 10 plates with thicknesses in the range $t$=$[0.180,\,1.080]$\,mm, measured using an optical microscope (VHX-950F, Keyence). Post-curing, we cut eight circular plates (\textbf{Figure~\ref{Fig5}}a) with diameters in the range $D_\mathrm{p}$=$[25,\,60]\,$mm, in increments of 5\,mm.

\textit{Magnetization:} Then, various steps are involved in the magnetization of the magnetic plate, as illustrated in the schematic diagrams in Figure~\ref{Fig5}(a)-(d). The cut circular h-MRE plate of diameter $D_\mathrm{p}$ from the fabricated plate (Figure~\ref{Fig5}a) does not possess any magnetic properties. Therefore, to magnetize the plate, we folded it symmetrically into a semicircle along $x$-axis (Figure~\ref{Fig5}b) and then into a quarter-circle along $y$-axis (Figure~\ref{Fig5}c). Third, we placed the folded quarter-circle plate in the impulse magnetizer (IM-K-010020-A, flux density $\approx 4.4\,$T, Magnet-Physik Dr. Steingroever GmbH) at an angle of $\alpha=45^{\circ}$ with respect to the positive $y$ axis, aligned with the edge of the quarter circle (Figure~\ref{Fig5}d). The magnetizer generates a magnetic field of strength $\mathbf{B}$, inducing a permanent magnetic dipole in the NdPrFeB particles.  Assuming a uniform particle dispersion within the polymer matrix, the magnetization magnitude computed from the volume average of the total magnetic moment of the individual particles is $M={\mu_{0}}^{-1} B^{\mathrm{r}}=$134.4~kAm$^{-1}$, where $B^{\mathrm{r}}$ is the residual magnetic flux density and $\mu_0$ the relative permeability of air. After unfolding, the magnetization pattern of the circular plate is four-fold symmetric (Figure~\ref{Fig5}e). In each of the plate's four quarters, $k=\{1,2,3,4\}$, the magnetization is expected to be $\mathbf{M}\approx M\hat{\mathbf{n}}_k$, pointing along the unit vector:
\begin{equation}
    \mathbf{\hat{n}}_k = -\cos \left(\alpha+(k-1) \pi/2\right) \hat{\mathbf{e}}_x - \sin \left(\alpha+(k-1) \pi/2\right) \hat{\mathbf{e}}_y, 
\label{eq:magnetization}
\end{equation}
where $\alpha=45^{\circ}$ is the orientation of the folded plate in the magnetizer (see Figure~\ref{Fig5}d). 

Empirically, we found that our chosen magnetization profile is more effective in inducing snap buckling than simpler patterns (\textit{e.g.}, uniform magnetization parallel or perpendicular to the plate mid-surface). Our choice aligns with the anti-symmetric profile selected in recent snap-buckling studies of h-MRE   beams~\cite{abbasi2022snap,tan2022dynamic}. We also acknowledge  that our choice of the magnetization profile is motivated by fabrication simplicity. However,  we recognize the need to conduct a more systematic exploration of other profiles~\cite{zhao2022topology} in future research.

Having fabricated and magnetized the plates, we proceeded to produce shallow shells through radial compression of the said plates. First, we fabricated two VPS-32 annuli to act as the clamped boundary of the shell, each with inner and outer diameters of $D=D_\mathrm{p}/2$ and $2D=D_\mathrm{p}$, respectively (Figure~\ref{Fig5}f). Next, we stretched these annuli using two rigid cylinders of diameter $D+\lambda D$ (Figure~\ref{Fig5}g), resulting in the radial pre-stretch of $\lambda=\Delta D/D$ (Figure~\ref{Fig5}g). The plate was then sandwiched and bonded between the  annuli using the same VPS material. After curing, the cylinders were removed, thereby relaxing the pre-stretched. Consequently, the plate buckled out-of-plane due to the in-plane ($x$-$y$) radial compression, yielding a shell, the braille dot (Figure~\ref{Fig5}h). The height, $h$, of this newly formed shell was measured using an optical profilometer (VR-3200, Keyence Corporation).

For the \textit{writing experiments} 
(Figure~\ref{Fig5}i), the sample designated for testing was placed within the region of the uniform magnetic field  produced by a set of Helmholtz coils~\cite{yan2021magneto, yan2021comprehensive, abbasi2022snap}. The sample was clamped between two acrylic (rigid) plates.
Gravitational effects were minimized by orienting the shell's snapping direction (along $\pm\hat{\mathbf{z}}$) perpendicularly to gravity (-g$\hat{\mathbf{y}}$). We determined the critical magnetic field $B^\mathrm{a}_\mathrm{c}$ needed for snap-buckling, thereby writing the desired state of the braille dot. To do so, we gradually increased the magnetic flux density by increasing the current $I$ in the coils, in increments of 0.05\,A over 20\,s intervals, until snap-through occurred~\cite{abbasi2022snap}.

For the \textit{reading experiments}, we used the apparatus shown in Figure~\ref{Fig5}(j). The specimen was mounted on an acrylic plate containing a hole to equalize the in-out differential pressure. The shell equator was clamped using its thick boundary annuli and mounted onto an acrylic plate using silicone glue. For the intender, we used a rigid acrylic disk of diameter $D_\mathrm{ind}=0.8D$, smaller than the inner diameter ($D$) of the thick annular boundary to prevent contact with the perimeter during indentation. The indentation displacement was imposed at a constant velocity, 0.06\,mms$^{-1}$. The reaction force, $f$, exerted on the indenter was measured by a load cell (2530-5N, Instron). We define the \textit{blocking force}, $F$, as the maximum of $f$, beyond which the shell undergoes snap-through buckling, altering the state of the dot.


\subsection{Finite element modeling simulations}
\label{FEM}

We conducted the FEM simulations using the commercial package ABAQUS/Standard, undertaking two distinct series of simulations with the same protocol but different parameters. First, we worked with the same parameters as the scaled-up model system described in Section~\ref{Experiments}. The objective was to validate the FEM simulations against experiments. For the second series of simulations, we shifted to the realistic dimensions of the braille dots discussed in Section~\ref{Braille_Design}. In this case, the plate thickness was varied in the range $t$=$[0.025,0.325]$\,mm (increments of 0.025\,mm), and the fabrication pre-stretch of the magnetic plate in the range $\lambda=[0.05-0.2]$ (increments of 0.025). 

The initially flat, circular magnetic plate was modeled as a three-dimensional solid body. Geometric nonlinearities were accounted for throughout the analysis. Similarly to the experiments, the plate was segmented into four quadrants, each with a magnetization oriented along $\mathbf{n}_k$ (cf Section~\ref{Experiments}, Equation~(\ref{eq:magnetization}) and Figure~\ref{Fig5}e). The magnetic plate was discretized using the user-defined 8-node brick element proposed by Zhao \textit{et al.}~\cite{zhao2019mechanics} for the modeling of hard-magnetic deformable solids under a uniform magnetic field. We conducted a convergence study  to determine the appropriate level of mesh discretization, resulting in 6 elements in the thickness direction, 60 elements along the diameter, and 200 elements circumstantially. Mechanical loads, both contact (indentation) and distributed (pressure), were applied via a dummy mesh of C3D8R solid elements sharing the same nodes as the user elements. The material was assumed to be an incompressible ($\nu\approx0.5$) Neo-Hookean solid with a bulk modulus 100 times higher 
than its shear modulus ($G=E/3$) and a Young's modulus of $E=1.76$~MPa.

In both the scaled-up and real-scale simulations, we explored various combinations of the parameters  ($t,\,D,\,\lambda$) to investigate (1) geometry of the fabricated dot (Figure~\ref{Fig1}b), (2) writing phase (Figure~\ref{Fig1}c), and (3) reading phase (Figure~\ref{Fig1}d), as specified next.

(1) \textit{Geometry:} To account for the possible emergence of higher-order modes during plate buckling, we simulated the entire magnetized plate without any symmetry assumptions (cf. Section~\ref{Experiments}). In order to break the symmetry on the $x-y$ plane and induce buckling, a small out-of-plane displacement ($0.01t$) was applied as an initial perturbation. Then, we specified the Dirichlet boundary condition on each boundary node and applied compression by imposing radially inward-directed displacements toward the center of the plate, a process that led to the formation of the shell. The extent of compression was set through $\lambda$. 

(2) \textit{Writing:} For the writing-phase simulations, having set the dot geometry in step (1), we subjected the raised dot (ON state) to a uniform external magnetic field, $B^\mathrm{a}=1\,$T. The field was applied with a slight misalignment of $1^{\circ}$ about the -$\hat{\mathbf{z}}$ direction to trigger asymmetrical buckling modes, thus providing a closer approximation of actual experimental conditions. The magnitude of the magnetic field was then increased  linearly in the range of [0,\,1]\,T. The magnetic field magnitude, $B^\mathrm{a}_\mathrm{c}$, needed to snap the dot to the second stable state, was determined from the magnetic-field increment at which the displacement of the shell pole exhibited a sudden jump. 

(3) \textit{Reading:} For the reading-phase simulations, the indentation was simulated using a rigid circular plate indenter that exerted controlled displacements and discretized using rigid elements. The contact between the indenter and the shell was assumed to be hard and frictionless, thereby preventing surface penetration and sliding. To quantify the mechanical force needed to induce snapping, the dots were subjected to a downward indentation load (along $-\hat{\mathbf{z}}$) until reversal occurred, and the blocking (maximum) force, $F$, was recorded.

(4) \textit{Pneumatic loading:} When simulating the real-scale braille dots, we also considered a constant pneumatic load, as discussed in Section~\ref{pressure}. This pressure loading served to widen the design space by stiffening the dot against snapping during the reading phase (indentation) and reducing the energy barrier during the writing phase. For each dot geometry, we first measured the critical pressure, $P_\mathrm{cr}$, required to snap the shell on its own, following the same procedure used to measure the critical magnetic field for snapping~\cite{yan2021magneto, abbasi2021probing}. A constant positive (or negative) pressure difference, within the range $|\Delta P| \in [0, P_\mathrm{cr})$ (in increments of $1\,$kPa) was applied normally to the surface of the shell before initiating the reading (or writing) simulation steps, respectively. Steps (2) and (3) described above were repeated under this constant pneumatic loading. Finally, the blocking force, $F$, and the critical magnetic field amplitude, $B^\mathrm{a}_\mathrm{c}$, were recorded for each pair of parameters ($\Delta P,\,t$). For these simulations, we set the fabrication pre-stretch to $\lambda=0.15$. \\

\medskip

\medskip
\noindent \textbf{Acknowledgements} \par 
We are grateful to Dong Yan for fruitful discussions and Naïs Coq for providing the illustration of Figure~\ref{Fig1}. A.A. acknowledges funding from the Federal Commission for Scholarships for Foreign Students (FCS) through a Swiss Government Excellence Scholarship (Grant No. 2019.0619). 

\medskip
\noindent \textbf{Conflict of Interest} \par 
The authors declare no conflict of interest.

\medskip
\noindent \textbf{Data Availability Statement} \par 
The data that support the findings of this study are available from the corresponding author upon reasonable request.
\medskip

\bibliographystyle{MSP}
\bibliography{Braille}


\end{document}


\centerline{\bf \large Designing a braille reader using the snap buckling of bistable magnetic shells} 
\vspace{5mm}
\centerline{\bf \large -- Supplemental Information -- }
\vspace{5mm}
\centerline{\bf \large Arefeh Abbasi, Tian Chen, Bastien F. G. Aymon, Pedro M. Reis}

\author{\centerline{\bf \large Arefeh Abbasi, Tian Chen, Bastien F. G. Aymon, Pedro M. Reis}}

\beginsupplement

\section*{\bf Dimensional data for the validation of the FEM simulations using scaled-up experiments}

\begin{figure}[h!]
    \centering
    \includegraphics[width=0.95\textwidth]{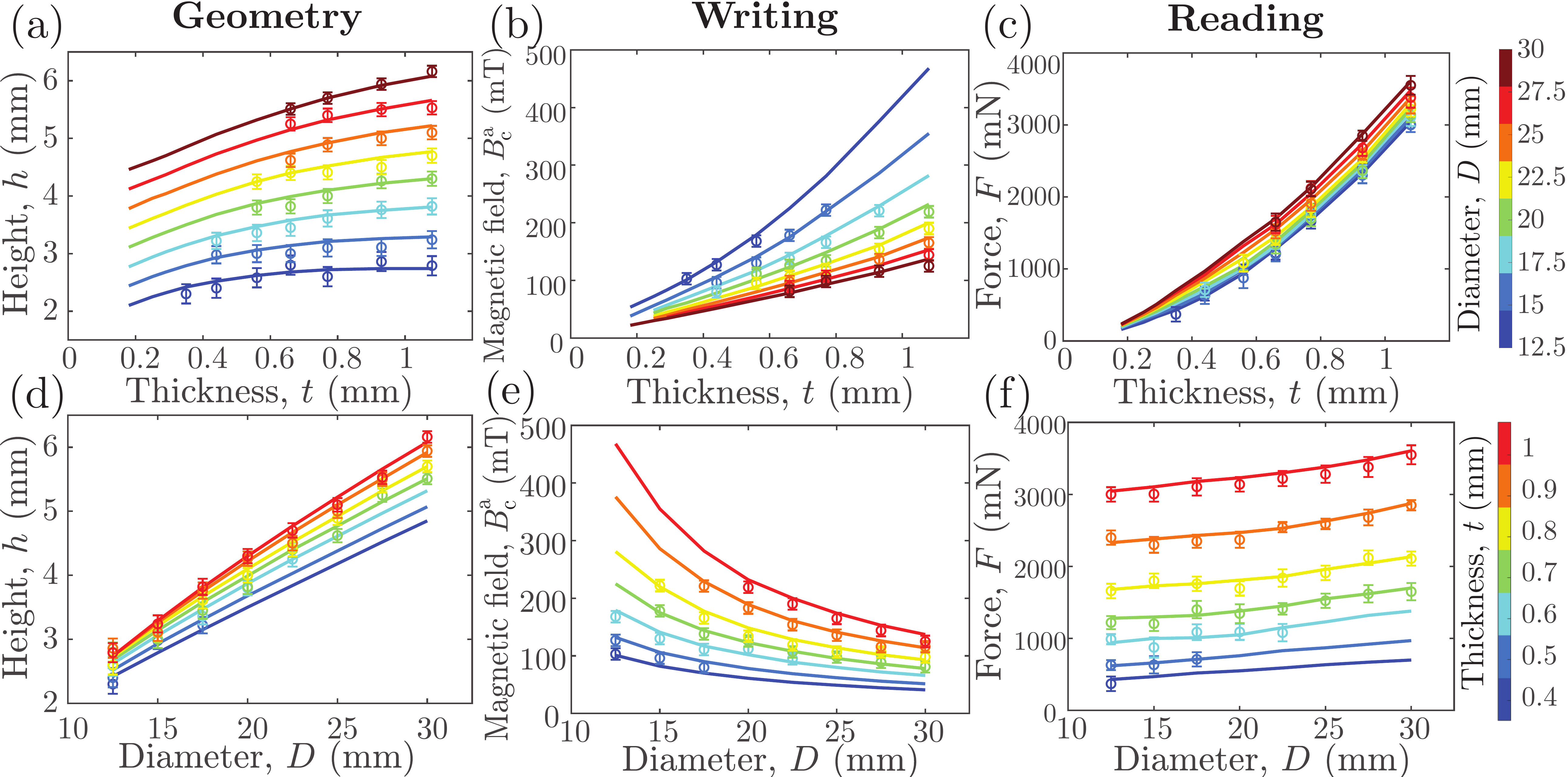}
    \caption{\textbf{FEM versus experiments data (dimensional):} (\textbf{a}) Dot height, $h$, (\textbf{b}) magnetic field, $B^\mathrm{a}_{c}$, and (\textbf{c}) indentation force, $F$, as a function of the thickness, $t$, for the diameter range of $D{\in}[12.5,\,30]$\,mm in increments of 2.5\,mm. (\textbf{d}) Dot height, $h$, (\textbf{e}) magnetic field, $B^\mathrm{a}_{c}$, and (\textbf{f}) indentation force, $F$, as a function of the diameter, $D$, for the thickness range of $t=\{0.18,\,0.25,\,0.29,\,0.35,\,0.44,\,0.56,\,0.66,\,0.77,\,0.93,\,1.08\}\,$mm. The solid lines represent the FEM results, and the symbols correspond to the experimental data. The error bars of the experimental data correspond to the standard deviation of different measurements on the same specimen.}
    \label{fig:rawdata}
 \end{figure}

Here, we present the dimensional version of the data presented in Figure 2 of the main manuscript for the geometry characterization (Figure~\ref{fig:rawdata}a,d), and both the writing (Figure~\ref{fig:rawdata}b,e) and reading (Figure~\ref{fig:rawdata}c,f) operational phases of the scaled-up dots. In \textbf{Figure~\ref{fig:rawdata}}(a,\,b,\,c), we plot the height $h$, the magnetic snapping load $B^\mathrm{a}_{c}$, and the maximum indentation force, $F$ as a function of the thickness $t$, respectively. In Figure~\ref{fig:rawdata}(d,\,e,\,f), we plot the $h$, $B^\mathrm{a}_{c}$, $F$ as a function of the diameter of the dot $D$, respectively. The diameter and thickness are color-coded in Figure~\ref{fig:rawdata}(a,\,b,\,c), and Figure~\ref{fig:rawdata}(d,\,e,\,f), respectively.

Figure~\ref{fig:rawdata}(a) and (d) show the height of the dot as a function of thickness $t$ and $D$, respectively. Increasing the thickness and diameter results in an increase in the dot height. When the diameter increases for a specific thickness value, the change in height becomes more significant compared to the increase in thickness. The FEM and experimental results are in good agreement, thus verifying the FEM framework for geometry.

In Figure~\ref{fig:rawdata}(b), and (e), we plot the critical magnetic field for snapping, $B^\mathrm{a}_{c}$, as a function of $t$, and $D$, respectively. We observed that as the thickness of the dot increases, so does the magnetic snapping load. However, increasing the diameter for each thickness has the opposite effect, as it leads to a decrease in the snapping load, which becomes more dominant in larger thicknesses. This is due to the fact that the slenderness ratio $D/t$ decreases. The FEM and experimental data match in this operational phase of the design as well.

In Figure~\ref{fig:rawdata}(c) and (f), which correspond to the reading phase, we plot the indentation force, $F$, is plotted as a function of $t$, and $D$, respectively. We find that increasing the thickness has a significant impact on the increase in force, $F$. However,  increasing the diameter only results in a slight increase in force. This is opposite to the relationship between $D$, $t$, and height. Again, the FEM and experiments are in excellent agreement.